%% file: acl2021.tex
\documentclass[11pt,a4paper]{article}
\usepackage[hyperref]{acl2021}
\usepackage{times}
\usepackage{latexsym}

\usepackage{microtype}

\usepackage{multirow}
\usepackage{amsmath}
\usepackage{amsthm}
\usepackage{amssymb}
\usepackage{graphicx}

\usepackage{microtype}

\aclfinalcopy 

\title{Contrastive Fine-tuning Improves Robustness for Neural Rankers}

\author{Xiaofei Ma \\
   \\
   \\\And
  Cicero Nogueira dos Santos \\
  AWS AI Labs \\
  \texttt{\{xiaofeim, cicnog, anarnld\}@amazon.com} \\
  \\\And
  Andrew O. Arnold \\
   \\
   \\}

\date{}

\begin{document}
\maketitle
\begin{abstract}
The performance of state-of-the-art neural rankers can deteriorate substantially when exposed to noisy inputs or applied to a new domain. In this paper, we present a novel method for fine-tuning neural rankers that can significantly improve their robustness to out-of-domain data and query perturbations. Specifically, a contrastive loss that compares data points in the representation space is combined with the standard ranking loss during fine-tuning. 
We use relevance labels to denote similar/dissimilar pairs, 
which allows the model to learn the underlying matching semantics across different query-document pairs and leads to improved robustness.
In experiments with four passage ranking datasets,
the proposed contrastive fine-tuning method obtains improvements on robustness to query reformulations, noise perturbations, and zero-shot transfer for both BERT and BART-based rankers. 
Additionally, our experiments show that contrastive fine-tuning outperforms data augmentation for robustifying neural rankers. 

\end{abstract}

\input{introduction.tex}

\input{method.tex}

\input{related_work.tex}

\input{exp_setup.tex}
\input{results.tex}

\section{Conclusion}
In this paper, we propose a novel method for fine-tuning neural rankers by combining contrastive loss with ranking loss. 
Using a semi-automatic approach, we created $6$ new versions of WikiQA test set to assess the robustness of our models to query reformulations and perturbations.
Our experimental results show that the proposed method improves ranker\textquotesingle s robustness to out-of-domain distributions, query reformulations, and perturbations. 
Comprehensive experiments and ablation studies were conducted to investigate the impact of some design choices as well as to confirm that the gains do not originate only from larger batch sizes.
Contrastive fine-tuning with generated data is more effective than data augmentation. As future work, we plan to evaluate the performance of other state-of-the-art contrastive loss functions and novel methods of aggregating multiple losses. 

\bibliographystyle{acl_natbib}
\bibliography{acl2021}

\input{appendix.tex}

\end{document}

%% file: introduction.tex
\section{Introduction}

\begin{figure*}[t]
\centering
    \includegraphics[width=0.95\textwidth]{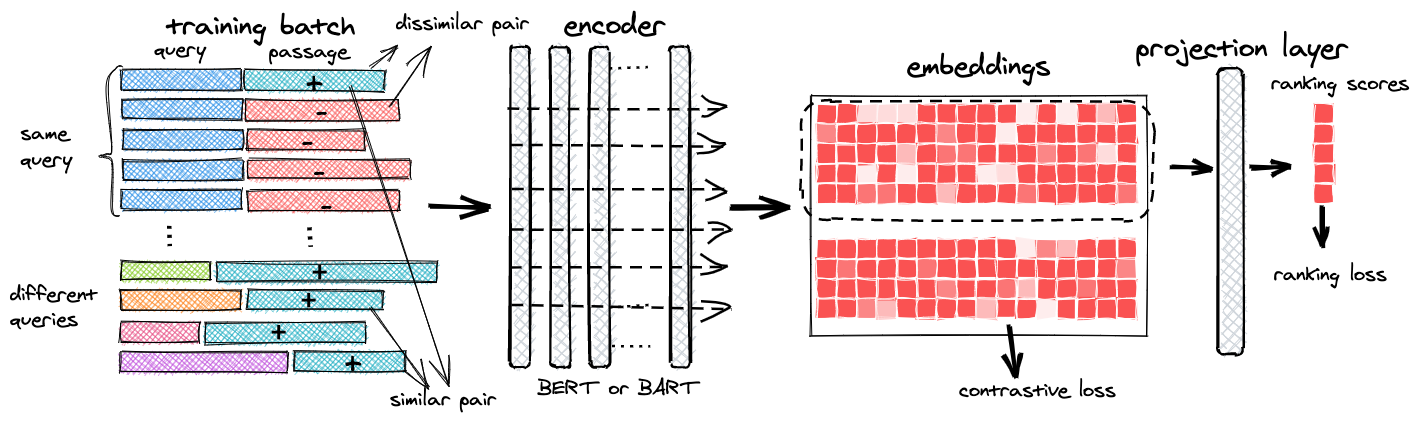}
    \caption{{\small Contrastive fine-tuning for neural rankers. During fine-tuning, a batch of positive and negative samples from different queries is fed into a neural encoder. The embeddings of query-document pairs from the same query are used to generate ranking scores, which are employed to compute the ranking loss. In parallel, the embeddings of all pairs are used to compute the contrastive loss.}}
    \label{fig:drawing}
\end{figure*}

Recent advances in neural language modeling have shifted the paradigm of natural language processing (NLP) towards a two-stage process: pre-training on a large amount of data with self-supervised tasks followed by fine-tuning on the target datasets with task-specific loss functions. 
Current state-of-the-art neural rankers for information retrieval fine-tune pre-trained language models using ranking losses on datasets containing examples of positive and negative query-document pairs.
While usually achieving good performance on in-domain test sets, neural rankers trained on large datasets can still exhibit poor transferability when tested in new domains, and suffer from robustness problems when exposed to various types of perturbations. For example, a neural ranker trained on a dataset with mostly natural language queries can perform badly when tested on keyword queries which are very common in information retrieval \citep{Bhatia2020}.


A considerable number of previous works have focused on domain adaptation to improve model \textquotesingle s overall transferability.
While domain adaptation approaches can help to address the out-of-domain robustness problem \citep{QiangYang2010, Zhang2019, Ma2019}, they rely on the availability of either labeled data or at least a target corpus which is usually not available at training time for a neural ranking model deployed in the wild.

The vulnerability of deep NLP models to various forms of adversarial attacks such as word-importance-based replacement \citep{Jin2020}, human-curated minimal perturbations \citep{Khashabi2020}, misspelling \citep{Sun2020}, grammatical errors \citep{Yin2020}, rule-based perturbations \citep{Si2020, Ribeiro2018} is well-documented in the literature \citep{EmmaZhang2019}. While various methods have been proposed to remediate model robustness issues in NLP, most of them are either task-specific \citep{Shah2019, Zhou2020, Gan2020, Wang2018}, requiring auxiliary tasks \citep{Zhou}, or relying on data augmentation \citep{Min2020, Kaushik2019, Cheng2019, Wei2019} which highly depends on the quality and diversity of the perturbed data.


An alternative strategy for optimizing machine learning models that has the potential to improve both out-of-domain generalization and robustness is contrastive learning.
Representations obtained under contrastive self-supervised settings have demonstrated improved robustness to out-of-domain distributions and image corruptions in computer vision tasks \citep{Hendrycks2019,radford2021}.
In contrastive learning, representations are learned by comparing among similar and dissimilar samples \citep{Le-Khac2020, Khosla2020, VanDenOord2018, Hjelm2018}. This is different from discriminative learning, where models learn a mapping of input samples to labels, and generative learning, where models reconstruct input samples. 
While several works have investigated contrastive learning for sentence classification \citep{Gunel2020}, sentence representation learning \citep{Wu2020}, and multi-modal representation learning \citep{radford2021} under either self-supervised or supervised settings, their potential for improving the robustness of neural rankers has not been explored yet.

In this paper, we propose a novel contrastive learning approach to fine-tune neural rankers and investigate its benefits for improving model robustness.
We focus on rankers that use single-tower architectures and are normally trained by optimizing a ranking loss that compares scores of positive and negative query-document pairs involving the same query. We propose to additionally use a contrastive loss that compares the distance between the representation of positive and negative pairs involving distinct queries (i.e, representations of positive pairs should be close in the latent space and distant to the representation of negative pairs, and vice-versa).
The goal of using this contrastive loss in addition to the ranking loss is to stimulate the model to learn the underlying matching semantics across different query-document pairs, which can potentially lead to improved robustness.


Our main contributions are as follows:
\begin{itemize}

\item We propose to combine contrastive loss with ranking loss during fine-tuning of neural ranking models and investigate its impact in improving model robustness and generalization.

\item Our experimental results using two language model-based neural rankers (BERT and BART) on four different datasets indicate that our proposed method improves upon standard ranking loss in zero-shot transfer across domains, leading to an increase of up to $9$ absolute points in Mean Average Precision (MAP).

\item We develop new datasets for evaluating the robustness of neural rankers. The datasets are based on WikiQA test set \cite{Yang2015} and were created semi-automatically. We plan to release these datasets upon acceptance of the paper.

\item We show that contrastive fine-tuned rankers are robust to 1) different types of query reformulations commonly seen in information retrieval (headline, paraphrase, and change of voice); and 2) query perturbations such as adding/removing punctuations, typos, and contractions/expansions.


\end{itemize}

%% file: method.tex
\section{Contrastive Representation Learning for Neural Ranking}
In neural ranking models, given an input query $q$ and a set of candidate documents $\{d_0, d_1, ..., d_n\}$, a neural network $h$ is used to create vector representations $\{h(q, d_1), ..., h(q, d_n)\}$, which are given to a function $s:\overrightarrow{x}\rightarrow \mathbb{R}$, that computes a score for each query-document pair, $\{s(h(q, d_1)), ..., s(h(q, d_n))\}$.
Normally $s$ performs a simple linear projection of the input embedding, and the training of neural ranking models consists
in optimizing a ranking loss that tries to enforce $s(h(q, d^+)) > s(h(q, d^-))$ for each training query $q$, where $d^+$ is a positive document for $q$ while $d^-$ is a negative one (See top part of Fig. \ref{fig:drawing}).


We propose to augment the training of neural rankers with the use of contrastive representation learning.
While ranking-based methods compute the loss with respect to the predicted scores, contrastive losses measure the distance/similarity between similar and dissimilar samples in the representation space. 
In our case,
the key idea consists in using a loss that compares the distance between the representation of query-document pairs, and enforces that positive pairs are close together in the latent space while being far apart from negative pairs, i.e., $D(h(q, d^+), h(q^{\prime}, d^{\prime+})) < D(h(q, d^+), h(q, d^-))$, where $q^{\prime}$ is either a variation of $q$ or a completely different query, and $d^{\prime+}$ is a positive document for $q^{\prime}$. Figure \ref{fig:drawing} illustrates our proposed approach, which is detailed in the remainder of this section.

\subsection{Ranking Loss}
Popular ranking losses include 1) the \emph{pairwise ranking loss}, in which the relevance information is given in the form of preferences between pairs of candidates, and 2) the \emph{listwise ranking loss} which directly optimizes a rank-based metric. 
In this work, we experiment with two pairwise ranking losses. The first one is the \textit{standard hinge loss} (SHL) defined on a triplet $(q, d^+, d^-)$ as follows:
\begin{eqnarray}
    \mathcal{L}_{SHL}(q, d^+, d^-; \theta) = \nonumber  \max\{0, \lambda - \\ s(h(q, d^+); \theta) + s(h(q, d^-); \theta)\}
\label{eq:SHL}
\end{eqnarray}

The other is a \textit{modified hinge loss} (MHL) function defined as:
\begin{eqnarray}
    \mathcal{L}_{MHL}(q, d^+, \{d^-_{i}\}; \theta) = \nonumber  \max\{0, \lambda - \\ s(h(q, d^+); \theta) + \max_{i}\{s(h(q, d^-_{i}); \theta)\}\}
\label{eq:MHL}
\end{eqnarray}
where $q$ is a query, $\lambda$ is the margin of the hinge loss, $d^+$ refer to the positive document. $d^-$ and \{$d^-_{i}\}$ refer to a negative document and the list of negative documents of the query $q$ within the same batch, respectively. $\theta$ includes the set of parameters of the network $h$ and the projection layer in $s$. Based on preliminary experiments, our modified ranking hinge loss generally performs better than the standard pairwise ranking hinge loss. Note that MHL loss has been used in previous work on passage ranking \cite{santosTXZ16}.

\subsection{Contrastive Loss}
For contrastive learning of representations, we employ the conceptually simple but widely adopted \emph{triplet margin loss} (\textbf{TML}) \citep{Weinberger, Chechik2010}, which has the following form:
\begin{eqnarray}
    \mathcal{L}_{TML}(a, k^+, k^-; \theta) = \nonumber\\ \max\{0, m + D(a, k^+; \theta) - D(a, k^-; \theta)\}
\label{eq:TML}
\end{eqnarray}
where $a$ is the anchor point, $k^+$ and $k^-$ are the similar and dissimilar samples with respect to the anchor point $a$. $m$ is the margin of the TML loss. In our neural ranking setting, an anchor point is the representation of a query-document pair. We use Euclidean or L2 distance $D$ in our experiment.
The contrastive loss can be applied to the representations from a variety of encoders $h(\cdot) \in R^d$. In this work, we explore contrastive fine-tuning for both BERT \cite{Devlin2018} and BART \cite{Lewis} models.

The key to effective contrastive learning is to design the \emph{notion of similarity} such that positive pairs may be very different in the input space yet semantically related.
In this work, we leverage the relevance label in the training data and consider as similar positive pairs $(q_i, d_i^+)$ and $(q_j, d_j^+)$ from different queries $i$ and $j$ in the same batch (as illustrated in Fig. \ref{fig:drawing}).
Our intuition is that,
by enforcing that positive pairs are close together in the embedding space and distant from negative pairs,
we make the scoring task easier.
Additionally, it allows the model to learn the underlying matching semantics across different query-document pairs, which leads to improved robustness.
We additionally conduct a brief experiment in Sec. \ref{subsec:data_augmentation} where we use paraphrases of the original query to generate similar pairs.



\subsection{Combined Loss} 
\label{sub:MOO}
Our final loss is a weighted average of the ranking loss $\mathcal{L}_{ranking}$ and the contrastive loss $\mathcal{L}_{contrastive}$:
\begin{eqnarray}
    \mathcal{L} = w_{1} \cdot \mathcal{L}_{ranking} + w_{2} \cdot \mathcal{L}_{contrastive}
\label{eq:overll}
\end{eqnarray}
The weights $w_{1}$ and $w_{2}$ are hyper-parameters that need to be determined. 
Our main experiments use a simple but effective combination method which consists in given equal weights to the ranking loss and contrastive loss.

%% file: related_work.tex
\section{Related Work}
Our work is related to the recent body of works that demonstrate contrastive and self-supervised approaches can improve model robustness and generalization.
\citet{Hendrycks2019} have shown that self-supervision increases image classifier\textquotesingle s robustness to adversarial examples, label corruption, and common input corruptions.
\citet{radford2021} have demonstrated that multi-modal contrastive learning can significantly improve the robustness of image classifiers to distribution shift.

In the NLP space, some recent works on sentence-level contrastive representation learning have shown its potential to improve robustness for classification \citep{Gunel2020} and semantic text similarity tasks \citep{Wu2020}.
There are two main distinctions between our work and these two papers: 1) they focus on classification and text similarity tasks, while we focus on ranking; and 2) while they rely on data augmentation approaches to define the notion of similarity, our approach mainly relies on document relevance information which is already present in the training data.

Our work is also related to recent work on neural retrieval that focus on hard negative mining to improve model performance \cite{gillick-etal-2019-learning,xiong2020approximate,karpukhin-etal-2020-dense,lu2020neural}.
The main differences between our work and this line of research are: 
1) while we leverage relevance information across different queries to create a notion of similarity, the focus on those papers are on finding hard negatives for each individual query in order to improve training efficiency. Hard negative mining can actually be used together with our method, as we show in Sec. \ref{subsec:hard_negative_mining}.
2) we focus in re-ranking models, which use single-tower model that create a single representation for a query-document pair. In contrast, neural retrieval models create separate representations for query and document.

%% file: exp_setup.tex
\section{Experimental Setup}
In this section, we describe the details of our experimental setup.

\subsection{Passage Ranking Datasets}
We test our method on four publicly available passage ranking/answer selection datasets that vary in size and domain. Passage ranking is an important task in information retrieval. It is often used to retrieve relevant content for open-domain question-answering systems \citep{Wang2020}.

\textbf{WikiQA} \citep{Yang2015} is a dataset of question and sentence pairs, collected and annotated for research on open-domain question answering. The questions are factoid and selected from Bing query logs. The answers are in the summary section of a linked Wikipedia page. The candidates are retrieved using Bing.

\textbf{WikiPassageQA} \citep{Cohen2018} is a benchmark collection for the research on non-factoid answer passage retrieval. The queries are created from Amazon Mechanical Turk over the top 863 Wikipedia documents from the Open Wikipedia Ranking.

\textbf{InsuranceQA} \citep{Feng2016} The question and answer pairs from this dataset are collected from the internet in the insurance domain. Each question has an answer pool of 500 candidates retrieved using SOLR.

\textbf{YahooQA} \citep{Tay2017} contains questions and answers from Yahoo! Answers website. The dataset is a subset of the Yahoo! Answers corpus from a 10/25/2007 dump. The questions are selected for their linguistic properties. For example, they all start with how $\{to \mid do \mid did \mid does \mid can \mid would \mid could \mid should\}$.

The statistics of the datasets are presented in Table \ref{tab:datasets}. All four datasets provide validation sets, which have size similar to the respective test sets.

\begin{table}[ht!]
\footnotesize
\scalebox{0.75}{
\begin{tabular}{llrr}
\hline
\textbf{Dataset} & \textbf{Domain} & \textbf{Train: \#Q (\#P/Q)} & \textbf{Test: \#Q (\#P/Q)}\\
\hline
WikiQA & Wikipedia & 873 (9) & 243 (9) \\
WikipassageQA & Wikipedia & 3,332 (58.3) & 416 (57.6) \\
InsuranceQA & insurance & 12,889 (500) & 2,000 (500) \\
YahooQA & community & 50,112 (5) & 6,283 (5) \\
\hline
\end{tabular}
}
\caption{{\small Dataset statistics. \#Q stands for \textit{number of questions} and \#P/Q is the \textit{average number of passages per question}}}\label{tab:datasets}
\end{table}

\subsection{Datasets for Robustness Assessment}
\label{sec:robustness_datasets}
In order to assess the robustness of our models to different types of query reformulations and query perturbations, we built robustness test datasets based on the original WikiQA test set.

We assessed query perturbations by leveraging CheckList \citep{Ribeiro2020} to construct three types of popular perturbations: \emph{adding/removing punctuation}, \emph{introducing typos} and \emph{changing of contraction form}.
For each query in WikiQA test set, we produce three new versions of the query, one for each perturbation type.

We assessed robustness to three types of query reformulations: \emph{paraphrase}, \emph{headline} and \emph{change of voice}.
We semi-automatically created the datasets for query reformulations in two steps:
(1) a pre-trained T5-base model \cite{Raffel2019} is fine-tuned on a combination of large public paraphrase datasets (Quora \footnote{https://www.quora.com/q/quoradata/First-Quora-Dataset-Release-Question-Pairs} and PAWS \citep{Zhang2019}) and human-curated query reformulations. 
The human-curated reformulations are based on the queries from the SQuAD 1.1 official dev set \footnote{https://rajpurkar.github.io/SQuAD-explorer/}. 
For each query in SQuAD 1.1 dev set, the annotators are asked to generate three new versions of the query (one for each reformulation type).
During fine-tuning and inference, we use control codes to instruct T5 on the type of reformulation to be generated. Note that this T5 model can be also used for the purpose of data augmentation, as shown in Sec. \ref{subsec:data_augmentation}.
(2) each query in WikiQA test set is processed by the fine-tuned T5 and three reformulations of the query are generated. All generated queries are post-processed in order to ensure it is grammatically correct and semantically equivalent to the original query. To ensure reliable evaluation, we did a round of human annotations to filter out low-quality generations. Examples of query reformulations are presented in Table \ref{tab:t5-gen}.

\begin{figure}[h!]
\includegraphics[width=0.49\textwidth]{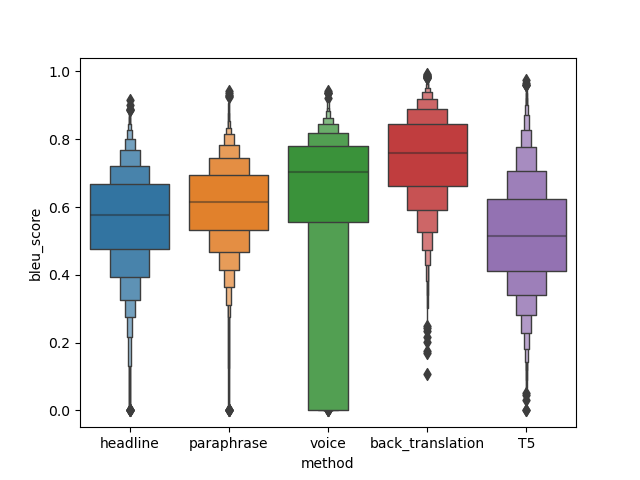}
\caption{Comparison of BLEU scores between original query and reformulations generated by human annotation, back translation and fine-tuned T5 model.}
\label{fig:bleu}
\end{figure}

We evaluate the lexical diversity of generated query reformulations by computing the BLEU scores between the original query and the reformulated query. The results of comparing four different generation methods are presented in Figure \ref{fig:bleu}. Our T5 generated queries overall exhibit higher diversity than human generated and back translation generated paraphrases. Note that the lower the BLEU score the higher the diversity.

\begin{table}[ht!]
\footnotesize
\scalebox{0.62}{
\begin{tabular}{ll}
\hline
\hline
\textbf{Query}
& \textit{What fueled the economy of early Vancouver?} \\
\hline
\textbf{Headline} & Factors that fuel Vancouver's economy \\
\textbf{Paraphrase} & What were some factors that stimulated the early Vancouver's economy? \\
\textbf{Chg of Voice} & The economy of early Vancouver was fueled by what? \\
\hline
\hline
\textbf{Query}
& \textit{How was the enlightenment shaped by science of the time period?} \\
\hline
\textbf{Headline} & type of science shaping enlightenment \\
\textbf{Paraphrase} & How was science of the time influenced the Enlightenment? \\
\textbf{Chg of Voice} & How did science of the time frame shape the Enlightenment? \\
\hline
\hline
\textbf{Query}
& \textit{How did South America gain independence from Spain and Portugal?} \\
\hline
\textbf{Headline} & Process of independence of South America from Spain and Portugal \\
\textbf{Paraphrase} & How did South America come to independence from Spain and Portugal? \\
\textbf{Chg of Voice} & How was independence from Spain and Portugal gained by South America? \\
\hline
\end{tabular}
}
\caption{\small Examples of T5 generated styled paraphrases} \label{tab:t5-gen}
\end{table}

\subsection{Neural Ranker Training}
We train neural rankers by fine-tuning two pre-trained language models: BERT and BART. For fine-tuning BERT, we use BERT-base model ($12$ layers, $110$M parameters) from Huggingface\textquotesingle s transformer codebase \cite{wolf2019HuggingFacesTS}.
Similar to the setup of sentence pair classification task in \citep{Devlin2018}, we concatenate the query sentence and the candidate passage together as a single input to the BERT encoder. 
We compute both the contrastive loss and ranking scores based on the [CLS] token embedding of the final hidden layer. 
For BART model fine-tuning, we use a BART-base model ($6$ layers encoder, $6$ layer decoder, $139$M parameters). We adopt the setting of BART for classification task in \citep{lewis-etal-2020-bart}. The concatenation of query text and passage text is fed into both the encoder and decoder and the last layer\textquotesingle s hidden state of the end decoder token is fed into a linear scorer. Similar to the [CLS] token in BERT, the embedding of the end token from the decoder is used as the representation of the complete input. For training with SHL, we sample triplets $(q, d^+, d^-)$ from different queries to form a single batch. For MHL training, a single batch consists of a positive passage $d^+$ and a list of negative passages $\{d^-_{i}\}$ from the same query $q$. We leverage the toolkit developed by \citet{musgrave2020pytorch} for contrastive loss calculation, and fine-tune the models for a maximum of $10$ epochs and adopt early stopping using the validation sets of each dataset. The hyper-parameters for fine-tuning neural rankers are listed in Appendix \ref{app:fine-tuning}.


%% file: results.tex
\section{Results and Discussion}
\label{sec:result}
\subsection{In-Domain Fine-tuning}
\label{sect:in-domain}
The results of in-domain fine-tuning of BERT and BART-based neural rankers on four passage ranking datasets are presented in table \ref{tab:in-domain results}. To ensure a fair comparison, all the hyper-parameters between the ranking and the contrastive settings are kept the same and equal weights between ranking loss and contrastive loss are used.
Our rankers produce state-of-the-art results when training with either of the two ranking-based losses \textbf{MHL} and \textbf{SHL}.
Adding contrastive loss (\textbf{TML}) slightly improves the in-domain performance for BERT-based rankers, and performs similarly as ranking loss for BART-based rankers.
Since our modified hinge loss (MHL) generally performs better than standard hinge loss (SHL), most of the results presented in the following sections are based on MHL, which corresponds to the setting illustrated in Figure \ref{fig:drawing}.

\begin{table*}[t!]
\centering
\footnotesize
\scalebox{0.79}{
\begin{tabular}{lcccccccccccc}
\hline
\textbf{Dataset $\rightarrow$} &  \multicolumn{3}{c}{\textbf{WikiQA}} & \multicolumn{3}{c}{\textbf{WikiPassageQA}} & \multicolumn{3}{c}{\textbf{InsuranceQA}} & \multicolumn{3}{c}{\textbf{YahooQA}} \\
\textbf{Model} & MAP & MRR & P@1 & MAP & MRR & P@1  & MAP & MRR & P@1  & MAP & MRR & P@1 \\
\hline
\multicolumn{13}{c}{\textbf{BERT}} \\
\hline
BERTSel-base \citeyearpar{Li2019}   
& 75.3 & 77.0 & - 
& - & - & - 
& - & - & - 
& 94.2 & 94.2 & - \\
BERT-PR-base \citeyearpar{Xu2019}  
& - & - & - 
& 73.5 & 80.9 & 70.2 
& 41.3 & 49.6 & 40.1 
& - & - & - \\
\hline
BERT-base-MHL (ours)
& 82.1 & 84.0 & 74.5
& 76.3 & 83.0 & 73.6
& 39.4 & 47.4 & 37.9
& 96.2 & 96.1 & 93.4 \\
BERT-base-MHL+TML (ours)  
& 83.8 & 85.8 & 77.4
& 76.9 & 83.1 & 73.6
& 41.1 & 49.6 & 39.5
& 96.1 & 96.1 & 93.4 \\
\hline
BERT-base-SHL (ours)
& 82.3 & 84.1 & 75.7
& 74.2 & 81.2 & 71.6
& 40.0 & 47.6 & 37.3
& 95.9 & 95.9 & 92.9 \\
BERT-base-SHL+TML (ours)
& 82.6 & 84.5 & 76.1
& 74.7 & 81.1 & 70.0
& 40.1 & 47.5 & 36.9
& 95.8 & 95.8 & 92.8 \\
\hline
\multicolumn{13}{c}{\textbf{BART}} \\
\hline
BART-base$_{LUL}$ \citeyearpar{NogueiradosSantos2020}
& 77.8 & 78.8 & 65.8 
& 73.8 & 81.3 & 71.9  
& 44.0 & 52.6 & 43.4
& 92.8 & 92.8 & 87.6 \\
BART-base$_{RLL}$ \citeyearpar{NogueiradosSantos2020}
& 77.5 & 79.2 & 65.4 
& 76.1 & 83.4 & 74.3  
& 42.2 & 50.3 & 40.8
& 96.1 & 96.1 & 93.4 \\
\hline
BART-base-MHL (ours)
& 85.8 & 87.4 & 78.2
& 77.8 & 85.3 & 77.6 
& 43.5 & 51.8 & 42.0
& 96.5 & 96.5 & 94.0 \\
BART-base-MHL+TML (ours)
& 84.6 & 86.1 & 75.7
& 77.4 & 84.5 & 76.2
& 43.4 & 51.9 & 42.4
& 96.5 & 96.5 & 93.9 \\
\hline
BART-base-SHL (ours)
& 82.4 & 83.9 & 73.3
& 75.9 & 83.1 & 73.6
& 42.9 & 51.2 & 41.6
& 96.0 & 96.0 & 93.1 \\
BART-base-SHL+TML (ours)
& 81.6 & 82.6 & 70.0
& 75.1 & 82.1 & 71.9
& 42.9 & 51.3 & 41.4
& 96.5 & 96.5 & 93.8 \\
\hline
\end{tabular}
}
\caption{{\small In-domain results of neural rankers trained on ranking loss vs contrastive loss.}} 
\label{tab:in-domain results}
\end{table*}

\begin{figure}[t]
\includegraphics[width=0.45\textwidth]{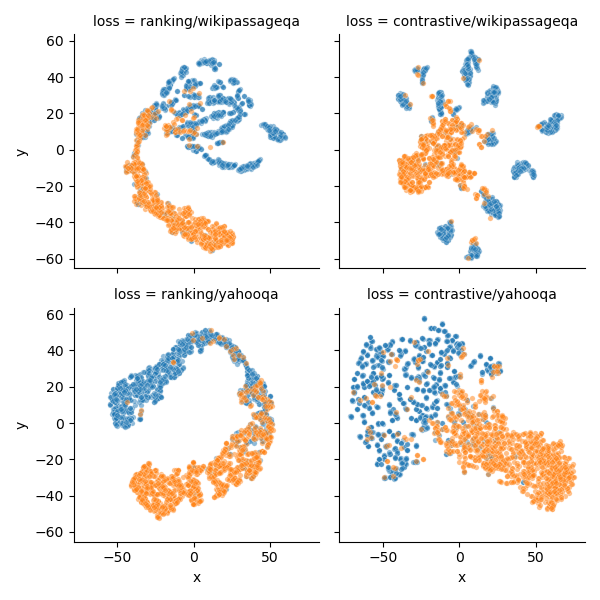}
\centering
\caption{t-SNE plot of representations from BERT-based rankers trained with either ranking or combined ranking and contrastive loss for samples from WikiPassageQA (top) and YahooQA (bottom) test sets. Positive samples are represented by orange color.}
\label{fig:tsne}
\end{figure}

To illustrate the effect of the contrastive loss on the representation space, we present the t-SNE plot of sample representations from the test set of two datasets WikiPassageQA and YahooQA in Figure \ref{fig:tsne}. The color in the figures represents the positive and negative labels of query-passage pairs. As we can see from the plots, adding contrastive loss enables further separation of the positive samples from the negative samples.

\subsection{Zero-Shot Transfer}
\label{ssec:zero-shot}

The zero-shot transfer performance of neural rankers reflects their robustness to out-of-domain distributions, which is a key property of neural rankers since they are usually deployed in the wild.
In Table \ref{tab:zero-shot results}, we show the results of applying the models trained in each one of the four datasets (source) and applied to the other three datasets (targets).
Overall, we see significant improvements in the zero-shot transferability of the model across all datasets when the neural ranker is trained using the combination of ranking and contrastive losses. The biggest improvement is from YahooQA $\rightarrow$ WikiPassageQA where we observe absolute $9$ points, $10.6$ points, and $11.8$ points improvement in MAP, MRR, and P@1, respectively. As expected, the transfer between datasets from similar domains tends to be better (e.g. WikiQA $\leftrightarrow$ WikiPassageQA) than that between dissimilar domains. 
Our intuition regarding the benefit of contrastive learning of representations to improve zero-shot transfer consists on the fact that, by using information from different queries and enforcing that positive pairs are close together in the embedding space and distant from negative pairs, the model ends up learning representations that are more general and therefore easier to transfer to new domains.



\begin{table*}[t!]
\centering
\footnotesize
\scalebox{0.78}{
\begin{tabular}{llcccccccccccc}
\hline
\  & \textbf{Target $\rightarrow$} &  \multicolumn{3}{c}{\textbf{WikiQA}} & \multicolumn{3}{c}{\textbf{WikiPassageQA}} & \multicolumn{3}{c}{\textbf{InsuranceQA}} & \multicolumn{3}{c}{\textbf{YahooQA}} \\
\textbf{Source Domain} & \textbf{Loss} & MAP & MRR & P@1 & MAP & MRR & P@1  & MAP & MRR & P@1  & MAP & MRR & P@1 \\
\hline
\multicolumn{13}{c}{\textbf{BERT}} \\
\hline
\textbf{WikiQA} & MHL   
& - & - & - 
& 58.3 & 66.4 & 51.7 
& 7.3 & 8.8 & 3.2 
& 49.4 & 49.3 & 26.7 \\
\textbf{WikiQA} & MHL+TML   
& - & - & - 
& 58.4 & 66.3 & 51.0 
& 11.6 & 14.9 & 7.5 
& 50.1 & 50.1 & 27.9 \\
\hline
\textbf{WikiPassageQA} & MHL
& 72.4 & 73.2 & 60.1 
& - & - & - 
& 30.1 & 36.8 & 26.9 
& 57.5 & 57.5 & 37.8 \\
\textbf{WikiPassageQA} & MHL+TML   
& 73.8 & 74.9 & 61.7 
& - & - & -
& 30.8 & 37.8 & 28.0 
& 59.5 & 59.5 & 40.1 \\
\hline
\textbf{InsuranceQA} & MHL   
& 59.6 & 60.4 & 41.6 
& 56.6 & 65.4 & 52.9 
& - & - & - 
& 69.2 & 69.2 & 52.0 \\
\textbf{InsuranceQA} & MHL+TML
& 61.1 & 62.3 & 45.3 
& 57.2 & 65.1 & 52.2 
& - & - & - 
& 78.0 & 78.0 & 64.3 \\
\hline
\textbf{YahooQA} & MHL   
& 35.3 & 36.3 & 16.5 
& 30.2 & 32.5 & 15.6 
& 3.8 & 4.4 & 0.9 
& - & - & - \\
\textbf{YahooQA} & MHL+TML   
& 38.7 & 39.5 & 18.1 
& 39.2 & 43.1 & 27.4 
& 5.5 & 6.2 & 1.7 
& - & - & - \\
\hline
\multicolumn{13}{c}{\textbf{BART}} \\
\hline
\textbf{WikiQA} & MHL   
& - & - & - 
& 52.4 & 58.9 & 41.4 
& 10.7 & 13.6 & 6.7 
& 51.8 & 51.8 & 30.0 \\
\textbf{WikiQA} & MHL+TML   
& - & - & -
& 58.6 & 67.1 & 51.9 
& 14.0 & 17.7 & 9.7 
& 52.4 & 52.4 & 30.9 \\
\hline
\textbf{WikiPassageQA} & MHL
& 74.6 & 75.9 & 63.0 
& - & - & - 
& 29.0 & 35.8 & 23.9 
& 62.6 & 62.6 & 44.7 \\
\textbf{WikiPassageQA} & MHL+TML   
& 76.5 & 78.1 & 66.3 
& - & - & - 
& 30.8 & 37.8 & 28.0 
& 63.8 & 63.8 & 45.7 \\
\hline
\textbf{InsuranceQA} & MHL   
& 64.9 & 66.1 & 50.2 
& 62.5 & 70.8 & 59.1 
& - & - & - 
& 69.1 & 69.1 & 51.6 \\
\textbf{InsuranceQA} & MHL+TML
& 65.5 & 66.3 & 50.6 
& 63.3 & 72.2 & 61.8 
& - & - & - 
& 71.7 & 71.7 & 55.4 \\
\hline
\textbf{YahooQA} & MHL   
& 34.1 & 35.4 & 15.6 
& 16.2 & 18.6 & 7.0 
& 2.2 & 2.7 & 0.6 
& - & - & - \\
\textbf{YahooQA} & MHL+TML   
& 35.3 & 36.2 & 17.3 
& 18.3 & 21.3 & 8.5 
& 2.5 & 3.2 & 0.6 
& - & - & - \\
\hline
\end{tabular}
}
\caption{{\small Results of zero-shot transfer for models trained on ranking loss vs. contrastive loss.}} \label{tab:zero-shot results}
\end{table*}

\subsection{Robustness to Query Perturbations}
In this section, we evaluate the model robustness to various types of reformulations and noisy transformations of the input queries.
The test sets used in the experiments are the $6$ variations of the WikiQA test set described in Sec. \ref{sec:robustness_datasets}.
We compare the results of using ranking loss (MHL) and the combination of ranking and contrastive loss (MHL+TML).
The robustness evaluation is are presented in Table \ref{tab:robustness_main}.
As shown in Table \ref{tab:robustness_main}, adding contrastive loss improves model robustness against all types of perturbations we tested.
We also conduct experiments by fine-tuning neural rankers on combined SHL loss and TML loss. The robustness evaluation of BERT-based rankers trained on SHL loss or combined SHL loss and TML loss are presented in Table \ref{tab:SHL}. Similar to the MHL case, the combined loss achieves a significant improvement in robustness than SHL loss only. More results on model robustness can be found in Appendix \ref{app:more}.




\begin{table*}[t!]
\centering
\footnotesize
\scalebox{0.75}{
\begin{tabular}{llcccccccccccc}
\hline
\  & \textbf{Reformulation $\rightarrow$} &  \multicolumn{2}{c}{\textbf{Headline}} & \multicolumn{2}{c}{\textbf{Paraphrase}} & \multicolumn{2}{c}{\textbf{Chg of Voice}} &
\multicolumn{2}{c}{\textbf{Punctuation}} &
\multicolumn{2}{c}{\textbf{Typo}} &
\multicolumn{2}{c}{\textbf{Contraction}} \\
\textbf{Training Data} & \textbf{Loss} & MAP & P@1 & MAP & P@1 & MAP & P@1 & MAP & P@1 & MAP & P@1 & MAP & P@1 \\
\hline
\multicolumn{14}{c}{\textbf{BERT}} \\
\hline
\textbf{WikiQA} & MHL   
& 75.3 & 64.9
& 79.4 & 69.3 
& 72.6 & 60.0
& 82.3 & 74.5
& 76.8 & 66.3 
& 82.6 & 77.6 \\
\textbf{WikiQA} & MHL+TML   
& 76.7 & 66.5 
& 81.8 & 73.4 
& 77.5 & 68.0 
& 83.0 & 75.3
& 79.4 & 71.3 
& 84.3 & 77.6 \\
\hline
\textbf{WikiPassageQA} & MHL
& 61.8 & 45.5
& 67.3 & 51.9 
& 64.2 & 48.0 
& 71.7 & 58.9
& 57.1 & 39.2 
& 74.7 & 64.3 \\
\textbf{WikiPassageQA} & MHL+TML 
& 63.3 & 47.5 
& 68.3 & 53.5 
& 68.9 & 56.0 
& 72.3 & 60.1
& 58.8 & 40.8 
& 74.9 & 64.3 \\
\hline
\multicolumn{14}{c}{\textbf{BART}} \\
\hline
\textbf{WikiQA} & MHL   
& 80.6 & 69.4
& 84.3 & 75.9 
& 76.8 & 64.0 
& 83.8 & 74.5
& 82.5 & 73.8 
& 86.3 & 79.6 \\
\textbf{WikiQA} & MHL+TML   
& 81.1 & 70.7
& 85.9 & 78.0 
& 79.2 & 72.0 
& 86.0 & 79.0
& 83.7 & 75.4 
& 86.5 & 77.6 \\
\hline
\textbf{WikiPassageQA} & MHL
& 72.0 & 58.7
& 75.3 & 62.7 
& 73.4 & 60.0 
& 74.6 & 62.6
& 70.0 & 55.8 
& 79.1 & 71.4 \\
\textbf{WikiPassageQA} & MHL+TML 
& 74.3 & 62.0
& 76.1 & 63.5 
& 76.0 & 64.0 
& 77.3 & 67.1
& 69.3 & 56.7 
& 79.6 & 70.4 \\
\hline
\end{tabular}
}
\caption{{\small Robustness to various types of query reformulations and perturbations for rankers with MHL loss.}} \label{tab:robustness_main}
\end{table*}

\begin{table*}[t!]
\centering
\footnotesize
\scalebox{0.75}{
\begin{tabular}{llcccccccccccc}
\hline
\  & \textbf{Reformulation $\rightarrow$} &  \multicolumn{2}{c}{\textbf{Headline}} & \multicolumn{2}{c}{\textbf{Paraphrase}} & \multicolumn{2}{c}{\textbf{Chg of Voice}} &
\multicolumn{2}{c}{\textbf{Punctuation}} &
\multicolumn{2}{c}{\textbf{Typo}} &
\multicolumn{2}{c}{\textbf{Contraction}} \\
\textbf{Training Data} & \textbf{Loss} & MAP & P@1 & MAP & P@1 & MAP & P@1 & MAP & P@1 & MAP & P@1 & MAP & P@1 \\
\hline
\textbf{WikiQA} & SHL   
& 73.1 & 62.0
& 79.6 & 71.0 
& 74.9 & 64.0 
& 81.0 & 72.4 
& 74.7 & 63.3
& 80.8 & 73.5 \\
\textbf{WikiQA} & SHL+TML   
& 75.2 & 64.5 
& 80.6 & 71.4
& 80.9 & 72.0
& 81.7 & 74.5
& 79.9 & 71.3
& 82.0 & 75.5 \\
\hline
\textbf{WikiPassageQA} & SHL
& 60.4 & 45.5
& 66.9 & 51.9 
& 54.6 & 40.0
& 68.9 & 54.7
& 55.1 & 36.7
& 69.6 & 55.1 \\
\textbf{WikiPassageQA} & SHL+TML 
& 62.5 & 46.3 
& 66.9 & 51.9 
& 62.7 & 48.0
& 68.9 & 54.7 
& 56.9 & 39.2
& 72.0 & 58.2 \\
\hline
\end{tabular}
}
\caption{{\small Robustness to various types of paraphrase for BERT-based rankers trained with SHL.}} \label{tab:SHL}
\end{table*}

\begin{table*}[t!]
\centering
\footnotesize
\scalebox{0.75}{
\begin{tabular}{lllccccccccc}
\hline
\  &  & \textbf{Reformulation $\rightarrow$} &  \multicolumn{3}{c}{\textbf{Headline}} & \multicolumn{3}{c}{\textbf{Paraphrase}} & \multicolumn{3}{c}{\textbf{Change of Voice}} \\
\textbf{Loss} & \textbf{Notion of Similarity} & \textbf{Training Data} & MAP & MRR & P@1 & MAP & MRR & P@1  & MAP & MRR & P@1 \\
\hline
\textbf{MHL} & - & WikiQA
& 75.3 & 76.9 & 64.9
& 79.4 & 80.9 & 69.3
& 72.6 & 73.6 & 60.0 \\
\textbf{MHL+TML} & relevance label & WikiQA
& 76.7 & 78.0 & 66.5
& 81.8 & 83.2 & 73.4
& 77.5 & 78.6 & 68.0 \\
\hline
\textbf{MHL} & - & WikiQA + headline   
& 76.1 & 77.6 & 63.6
& 78.2 & 79.7 & 66.4 
& 73.8 & 75.2 & 60.0 \\
\textbf{MHL+TML} & query reformulations & WikiQA + headline & 77.1 & 78.6 & 65.7 
& 80.2 & 81.4 & 71.0 
& 81.2 & 83.7 & 76.0  \\
\hline
\textbf{MHL} & - & WikiQA + paraphrase
& 70.0 & 71.0 & 54.6
& 81.3 & 82.6 & 71.8
& 78.9 & 80.3 & 68.0 \\
\textbf{MHL+TML} & query reformulations & WikiQA + paraphrase 
& 75.1 & 76.4 & 63.2 
& 81.2 & 82.4 & 70.5 
& 80.3 & 83.1 & 76.0 \\
\hline
\textbf{MHL} & - & WikiQA + chg voice   
& 73.6 & 75.0 & 60.7 
& 82.8 & 83.9 & 73.4
& 81.2 & 84.5 & 76.0 \\
\textbf{MHL+TML} & query reformulations & WikiQA + chg voice
& 75.3 & 76.6 & 63.2
& 83.0 & 84.3 & 73.4 
& 83.3 & 88.5 & 80.0 \\
\hline
\end{tabular}
}
\caption{{\small Comparison with data argumentation.}} \label{tab:argumentation}
\end{table*}

\subsection{Comparison with Data Augmentation}
\label{subsec:data_augmentation}
One of the traditional approaches for improving the robustness of machine learning models is to augment the training data with noisy data.
In this section, we compare our contrastive fine-tuning method with a data augmentation approach in which automatically generated query reformulations are added to the training data. For each query in the training set, we use our fine-tuned T5 model to generate 5 new queries of each reformulation type (headline, paraphrase, change of voice). Effectively, we increase the training set and the number of training steps by a factor of 5 for each reformulation type. 
Since our proposed training approach is general and can be used with any dataset, we also experiment with data augmentation combined with contrastive fine-tuning. 
When performing contrastive fine-tuning, we use the 5 reformulations of each query to create similar pairs (the notion of similarity == query formulation) in each batch, which essentially keeps the number of training steps the same as when training with the original dataset.
The results on data augmentation for BERT-based neural rankers are presented in Table \ref{tab:argumentation}. For rows with MHL loss, we argument the training data with paraphrased queries and train the model on a combined dataset using MHL loss only. Note this is the standard way to do data argumentation training. For rows using MHL+TML loss, we pair each query in the batch with its paraphrased query for contrastive loss calculation. The model is trained on combined MHL+TML loss. For both methods, we expose the model with the same amount of paraphrased training samples. As we can see from the table, augmenting the training data with a similar type of query reformulation can improve the robustness of the model against that particular type of reformulation.
However, it is not as effective in improving the robustness against other reformulation types. 
On the other hand, contrastive fine-tuning, even trained with a single type of query reformulation can generally improve the model robustness against the other two types. 
Furthermore, contrastive fine-tuning achieves this with significantly less ($4\times$ less) training time even after considering the additional computation of the contrastive loss calculation. Essentially, the experimental results indicate that using paraphrased training samples to perform contrastive learning is both effective (produces more robust rankers) as well as efficient (faster to train since augmented data is used in parallel, i.e. same batch as original data) than using regular data augmentation.

\subsection{Comparison with Ranking Loss using same Batch Size}
When training with MHL plus contrastive loss we effectively increase the batch size because we need to augment the training batch with additional positive samples from different queries.
To check if the improvements achieved by our approach are due to the increase of training batch size only, we perform an ablation study where we compare the performance of models trained with a ranking loss but with the same batch size as the contrastive fine-tuning setting. 
The comparison results are presented in Table \ref{tab:batch_size}. 
As we can see in the table, for WikiQA datase, increasing the batch size helps the performance of in-domain and some of the robustness test sets. The contrastive setting still outperforms the ranking setting in all the test categories. Increasing the batch size of MHL is not always beneficial. We see big degradation on the WikiPassageQA dataset. On the other hand, we observed consistent improvement when the model is trained with contrastive loss.

\section{Ablation Study}
We present ablation experiments that check the impact of the number of positive samples per batch and the use of hard negative mining.
Additionally, in Appendix \ref{app:moo}, we also present a preliminary experiment on formulating our combined loss (Equation \ref{eq:overll}) as a multi-objective optimization (MOO).


\begin{table*}[t!]
\centering
\footnotesize
\scalebox{0.9}{
\begin{tabular}{lllcccccccc}
\hline
\  & & \textbf{Test Set $\rightarrow$} & \multicolumn{2}{c}{\textbf{In-domain}} & \multicolumn{2}{c}{\textbf{Headline}} & \multicolumn{2}{c}{\textbf{Paraphrase}} & \multicolumn{2}{c}{\textbf{Chg of Voice}} \\
\textbf{Loss} & \textbf{Batch Size} & \textbf{Training Set} & MAP & P@1 & MAP & P@1 & MAP & P@1  & MAP & P@1 \\
\hline
\textbf{MHL} & 16 & WikiQA   
& 82.1 & 74.5   
& 75.3 & 64.9 
& 79.4 & 69.3  
& 72.6 & 60.0 \\
\textbf{MHL} & 31 & WikiQA   
& 83.0 & 76.1 
& 74.9 & 64.1 
& 81.0 & 71.8
& 77.2 & 68.0 \\
\textbf{MHL+TML} & 31 & WikiQA 
& 83.8 & 77.4
& 76.7 & 66.5 
& 81.8 & 73.4  
& 77.5 & 68.0 \\
\hline
\textbf{MHL} & 16 & WikiPassageQA
& 76.3 & 73.6
& 61.8 & 45.5
& 67.3 & 51.9
& 64.2 & 48.0 \\
\textbf{MHL} & 31 & WikiPassageQA
& 76.2 & 73.8
& 58.7 & 40.5
& 64.4 & 47.7
& 57.1 & 40.0 \\
\textbf{MHL+TML} & 31 & WikiPassageQA 
& 76.9 & 73.6
& 63.3 & 47.5 
& 68.3 & 53.5
& 68.9 & 56.0 \\
\hline
\end{tabular}
}
\caption{{\small Comparison of ranking loss and contrastive loss with same batch size for BERT-based rankers.}} \label{tab:batch_size}
\end{table*}

\begin{table*}[t!]
\centering
\footnotesize
\scalebox{0.88}{
\begin{tabular}{lcccccccccccc}
\hline
\textbf{Target $\rightarrow$} &  \multicolumn{3}{c}{\textbf{WikiQA}} & \multicolumn{3}{c}{\textbf{WikiPassageQA}} & \multicolumn{3}{c}{\textbf{InsuranceQA}} & \multicolumn{3}{c}{\textbf{YahooQA}} \\
\textbf{\# Pos / Batch} & MAP & MRR & P@1 & MAP & MRR & P@1  & MAP & MRR & P@1  & MAP & MRR & P@1 \\
\hline
2
& \underline{82.4} & \underline{84.1} & \underline{74.9} 
& 58.1 & 66.4 & 51.7
& 12.8 & 16.3 & 8.0
& 46.4 & 46.3 & 24.2 \\
4
& \underline{83.4} & \underline{85.2} & \underline{76.1} 
& 58.1 & 66.1 & 51.0
& 12.1 & 15.3 & 7.0 
& 47.9 & 47.8 & 26.0 \\
8  
& \underline{83.5} & \underline{85.3} & \underline{75.7} 
& 59.9 & 67.4 & 52.9 
& 12.1 & 15.2 & 7.1 
& 48.7 & 48.7 & 26.8 \\
16 
& \underline{83.8} & \underline{85.8} & \underline{77.4} 
& 59.4 & 66.3 & 51.0 
& 11.6 & 14.5 & 6.4 
& 50.1 & 50.1 & 27.9 \\
32
& \underline{83.3} & \underline{85.2} & \underline{76.1}  
& 59.0 & 65.9 & 49.3 
& 10.4 & 12.9 & 4.9 
& 49.9 & 49.9 & 27.6 \\
64
& \underline{82.5} & \underline{84.2} & \underline{74.5}
& 61.4 & 69.1 & 55.1
& 13.0 & 16.7 & 8.2 
& 49.2 & 49.2 & 26.9 \\
\hline
\end{tabular}
}
\caption{{\small Effect of number of positive pairs per batch. \underline{Underlined} cells indicate in-domain results.}} 
\label{tab:positive_number}
\end{table*}

\begin{table*}[t!]
\centering
\footnotesize
\scalebox{0.86}{
\begin{tabular}{lcccccccccccc}
\hline
\ \textbf{Target $\rightarrow$} &  \multicolumn{3}{c}{\textbf{WikiQA}} & \multicolumn{3}{c}{\textbf{WikiPassageQA}} & \multicolumn{3}{c}{\textbf{InsuranceQA}} & \multicolumn{3}{c}{\textbf{YahooQA}} \\
\textbf{Mining Method} & MAP & MRR & P@1 & MAP & MRR & P@1  & MAP & MRR & P@1  & MAP & MRR & P@1 \\
\hline
\multicolumn{13}{c}{\textbf{BERT}} \\
\hline
No Mining   
& \underline{83.8} & \underline{85.8} & \underline{77.4} 
& 59.4 & 66.3 & 51.0
& 11.6 & 14.9 & 7.5
& 50.1 & 50.1 & 27.9 \\
Angular   
& \underline{83.8} & \underline{85.6} & \underline{76.1} 
& 59.1 & 66.2 & 51.0 
& 6.7 & 8.1 & 3.1 
& 49.7 & 49.6 & 27.4 \\
BatchHard   
& \underline{82.3} & \underline{84.0} & \underline{73.7} 
& 63.0 & 70.7 & 57.0
& 10.7 & 13.9 & 6.4 
& 52.0 & 52.0 & 30.2 \\
TripletMargin   
& \underline{83.6} & \underline{85.6} & \underline{77.0} 
& 59.4 & 66.4 & 51.0 
& 7.2 & 8.7 & 3.2 
& 50.1 & 50.1 & 27.8 \\
\hline
\multicolumn{13}{c}{\textbf{BART}} \\
\hline
No Mining   
& \underline{84.6} & \underline{86.1} & \underline{75.8} 
& 58.6 & 67.1 & 51.9 
& 14.0 & 17.7 & 9.7 
& 52.4 & 52.3 & 30.9 \\
Angular   
& \underline{84.5} & \underline{86.1} & \underline{76.1} 
& 59.9 & 67.9 & 53.1 
& 16.0 & 20.0 & 11.3  
& 56.4 & 56.4 & 35.6 \\
BatchHard   
& \underline{85.4} & \underline{86.7} & \underline{77.0} 
& 57.8 & 66.2 & 50.5 
& 12.5 & 15.8 & 7.9 
& 55.3 & 55.3 & 33.9 \\
TripletMargin   
& \underline{85.9} & \underline{87.5} & \underline{78.6} 
& 61.2 & 69.6 & 55.3 
& 16.0 & 19.9 & 11.0 
& 56.6 & 56.5 & 35.8 \\
\hline
\end{tabular}
}
\caption{Effect of hard negative mining. \underline{Underlined} cells indicate in-domain results.} 
\label{tab:hard negative}
\end{table*}

\subsection{Effect of Number of Positive Samples Per Batch}
The number of positive samples within a single batch determines the total number of potential triples constructed. In this section, we vary the number of positive samples within a batch and evaluate its effect on the model performance. The results are presented in Table \ref{tab:positive_number}. 
As expected, the model performance benefits by increasing the number of positives in a batch.
As shown in Table \ref{tab:positive_number}, although not strictly monotonically, both the in-domain performance and zero-shot transfer performance improve with the number of positive pairs.

\subsection{Effect of Hard Negative Mining} \label{subsec:hard_negative_mining}
In a batch of N samples, there are $O(N^3)$ possible triplets, many of which are not very helpful to model convergence (e.g triplets where $D(a, k^+) >> D(a, k^-)$). It's important to construct only the most important triplets. Many works have discussed the benefit of hard negative mining techniques that produce useful gradients and help the models converge quickly. In this section, we explore the effect of three hard negative mining methods that are compatible with TML: Angular miner \citep{Wang2017} (output triplets that form an angle greater than a threshold), BatchHard \citep{Hermans2017} (produce a single triplet for each anchor point consisting of the hardest positive and hardest negative samples), and TripletMargin (only output a triplet when the difference between the anchor-positive distance and the anchor-negative distance is smaller than a margin). The results of hard negative mining on models trained on WikiQA dataset are presented in Table \ref{tab:hard negative}, in which we evaluate both the in-domain and zero-shot performance of the rankers. As we can see from the results, hard negative mining can further improve the transferability of of both BERT-based ranker and BART-based ranker. In particular, BatchHard outperforms other mining methods and improve the overall performance significantly for BERT-based rankers while TripletMargin is more effective for BART-based rankers. We believe there is still a margin for improvement if the hyper-parameters of the miners are properly tuned.

%% file: appendix.tex
\appendix

\section{Details of Neural Ranker Fine-tuning}
\label{app:fine-tuning}

The fine-tuning of neural rankers is conducted on an AWS EC2 P3 machine. Important hyper-parameters of fine-tuning for each model-dataset combination are listed in Table \ref{tab:hyper-parameters}. 

\begin{table*}[t!]
\centering
\footnotesize
\scalebox{0.74}{
\begin{tabular}{llcccccc}
\textbf{Training Data} & \textbf{Loss} & \textbf{Learning Rate} & \textbf{Positives / Batch} & \textbf{Negatives / Batch} & \textbf{Gradient Accumulation} & \textbf{Block Size} & \textbf{Hinge Loss Margin} \\
\hline
\multicolumn{8}{c}{\textbf{BERT}} \\
\hline
\textbf{WikiQA} & MHL   
& 5e-6 & 1 & 15 & 8 & 256 & 2 \\
\textbf{WikiQA} & MHL+TML   
& 5e-6 & 16 & 15 & 8 & 256 & 2  \\
\hline
\textbf{WikiPassageQA} & MHL
& 1e-5 & 1 & 15 & 8 & 256 & 2  \\
\textbf{WikiPassageQA} & MHL+TML   
& 1e-5 & 16 & 15 & 8 & 256 & 2  \\
\hline
\textbf{InsuranceQA} & MHL 
& 5e-5 & 1 & 15 & 16 & 256 & 2  \\
\textbf{InsuranceQA} & MHL+TML
& 5e-5 & 16 & 15 & 16 & 256 & 2 \\
\hline
\textbf{YahooQA} & MHL   
& 1e-5 & 1 & 4 & 8 & 256 & 2  \\
\textbf{YahooQA} & 
MHL+TML
& 1e-5 & 16 & 4 & 8 & 256 & 2  \\
\hline
\textbf{WikiQA} & SHL   
& 5e-6 & 15 & 15 & 8 & 256 & 2 \\
\textbf{WikiQA} & SHL+TML   
& 5e-6 & 15 & 15 & 8 & 256 & 2  \\
\hline
\textbf{WikiPassageQA} & SHL
& 1e-5 & 15 & 15 & 8 & 256 & 2  \\
\textbf{WikiPassageQA} & SHL+TML   
& 1e-5 & 15 & 15 & 8 & 256 & 2  \\
\hline
\textbf{InsuranceQA} & SHL 
& 5e-5 & 15 & 15 & 16 & 256 & 2  \\
\textbf{InsuranceQA} & SHL+TML
& 5e-5 & 15 & 15 & 16 & 256 & 2 \\
\hline
\textbf{YahooQA} & SHL   
& 1e-5 & 4 & 4 & 8 & 256 & 2  \\
\textbf{YahooQA} & 
SHL+TML
& 1e-5 & 4 & 4 & 8 & 256 & 2  \\
\hline
\multicolumn{8}{c}{\textbf{BART}} \\
\hline
\textbf{WikiQA} & MHL   
& 5e-6 & 1 & 15 & 8 & 256 & 2 \\
\textbf{WikiQA} & MHL+TML   
& 5e-6 & 16 & 15 & 8 & 256 & 2  \\
\hline
\textbf{WikiPassageQA} & MHL
& 1e-5 & 1 & 15 & 8 & 256 & 2  \\
\textbf{WikiPassageQA} & MHL+TML   
& 1e-5 & 16 & 15 & 8 & 256 & 2  \\
\hline
\textbf{InsuranceQA} & MHL 
& 5e-5 & 1 & 15 & 16 & 256 & 2  \\
\textbf{InsuranceQA} & MHL+TML
& 5e-5 & 16 & 15 & 16 & 256 & 2 \\
\hline
\textbf{YahooQA} & MHL   
& 1e-5 & 1 & 4 & 8 & 256 & 2  \\
\textbf{YahooQA} & 
MHL+TML
& 1e-5 & 16 & 4 & 8 & 256 & 2  \\
\hline
\textbf{WikiQA} & SHL   
& 5e-6 & 15 & 15 & 8 & 256 & 2 \\
\textbf{WikiQA} & SHL+TML   
& 5e-6 & 15 & 15 & 8 & 256 & 2  \\
\hline
\textbf{WikiPassageQA} & SHL
& 1e-5 & 15 & 15 & 8 & 256 & 2  \\
\textbf{WikiPassageQA} & SHL+TML   
& 1e-5 & 15 & 15 & 8 & 256 & 2  \\
\hline
\textbf{InsuranceQA} & SHL 
& 5e-5 & 15 & 15 & 16 & 256 & 2  \\
\textbf{InsuranceQA} & SHL+TML
& 5e-5 & 15 & 15 & 16 & 256 & 2 \\
\hline
\textbf{YahooQA} & SHL   
& 1e-5 & 4 & 4 & 8 & 256 & 2  \\
\textbf{YahooQA} & 
SHL+TML
& 1e-5 & 4 & 4 & 8 & 256 & 2  \\
\hline
\end{tabular}
}
\caption{{\small Hyper-parameters for neural ranker fine-tuning.}} \label{tab:hyper-parameters}
\end{table*}

\section{More Results of Robustness Against Query Perturbations}
\label{app:more}

In this section, we present more results of model robustness evaluation for neural rankers trained on InsuranceQA and YahooQA datasets. The results are shown in Table \ref{tab:more}.

\begin{table*}[t!]
\centering
\footnotesize
\scalebox{0.78}{
\begin{tabular}{llccccccccccccc}
\hline
\  & \textbf{Reformulation $\rightarrow$} &  \multicolumn{2}{c}{\textbf{Headline}} & \multicolumn{2}{c}{\textbf{Paraphrase}} & \multicolumn{2}{c}{\textbf{Chg of Voice}} &
\multicolumn{2}{c}{\textbf{Punctuation}} &
\multicolumn{2}{c}{\textbf{Typo}} &
\multicolumn{2}{c}{\textbf{Contraction}} & \\
\textbf{Training Data} & \textbf{Loss} & MAP & P@1 & MAP & P@1 & MAP & P@1 & MAP & P@1 & MAP & P@1 & MAP & P@1 & \textbf{Avg MAP} \\
\hline
\multicolumn{15}{c}{\textbf{BERT}} \\
\hline
\textbf{InsuranceQA} & MHL   
& 53.2 & 34.7 
& 57.4 & 39.8 
& 44.4 & 24.0 
& 59.2 & 41.2
& 42.8 & 23.4 
& 60.4 & 42.9 & 52.9 \\
\textbf{InsuranceQA} & MHL+TML
& 54.9 & 35.1
& 58.9 & 41.5 
& 47.9 & 28.0 
& 60.6 & 44.4
& 44.4 & 25.0 
& 60.0 & 42.9 & \textbf{54.5} \\
\hline
\textbf{YahooQA} & MHL   
& 40.5 & 22.3
& 34.0 & 14.1 
& 28.8 & 4.0 
& 36.7 & 16.9
& 34.0 & 14.6 
& 38.3 & 16.3 & 35.4 \\
\textbf{YahooQA} & MHL+TML 
& 41.0 & 21.9
& 35.5 & 14.5 
& 29.4 & 8.0  
& 40.1 & 20.2
& 35.5 & 15.0 
& 41.6 & 20.4 & \textbf{37.2} \\
\hline
\multicolumn{15}{c}{\textbf{BART}} \\
\hline
\textbf{InsuranceQA} & MHL 
& 61.2 & 46.3 
& 64.2 & 50.2 
& 52.6 & 32.0 
& 63.6 & 48.2
& 54.7 & 37.9 
& 68.6 & 54.1 & 60.8 \\
\textbf{InsuranceQA} & MHL+TML
& 63.4 & 50.8
& 65.0 & 50.6 
& 58.5 & 44.0 
& 64.7 & 50.2
& 54.6 & 37.5 
& 68.8 & 56.1 & \textbf{62.5} \\
\hline
\textbf{YahooQA} & MHL
& 35.7 & 17.4
& 32.9 & 13.7 
& 29.7 & 8.0 
& 34.2 & 15.2
& 33.6 & 15.8 
& 36.5 & 15.3 & 33.8 \\
\textbf{YahooQA} & MHL+TML   
& 37.2 & 18.2
& 34.4 & 15.8 
& 30.7 & 8.0 
& 35.2 & 16.9
& 34.6 & 17.5 
& 37.3 & 17.4 & \textbf{34.9} \\
\hline
\end{tabular}
}
\caption{{\small Additional restuls of robustness to various types of paraphrase.}} \label{tab:more}
\end{table*}



\section{Fine-tuning as Multi-objective Optimization}
\label{app:moo}
We performed a preliminary experiment on formulating equation \ref{eq:overll} as a multi-objective optimization (MOO) problem in which optimizing both $\mathcal{L}_{ranking}$ and $\mathcal{L}_{contrastive}$ are two objectives of the task. We adopt a \emph{dynamic weighted aggregation} (DWA) method \citep{Jin2001, Jin2004} which is both effective and computationally efficient. In DWA, the weights of the two loss terms are changed gradually according to the following equations:
\begin{eqnarray}
    w_{1}(t) = |\sin{2\pi t /\ F}|\\ 
    w_{2}(t) = 1 - w_{1}(t)
\label{eq:dynamic}
\end{eqnarray}
where $t$ is the iteration number. It is noticed that $w_{1}(t)$ changes from $0$ to $1$ periodically. The change frequency can be adjusted by $F$.

\begin{figure}[t]
\includegraphics[width=0.45\textwidth]{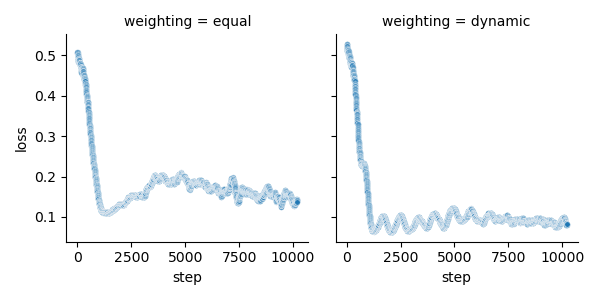}
\centering
\caption{Training contrastive loss for neural ranker trained on WikiQA.}
\label{fig:contrative_loss}
\end{figure}

Figure \ref{fig:contrative_loss} shows the evolution of the contrastive loss during fine-tuning of the BART-based ranker on the WikiQA dataset. As can be seen from the plot, adopting the MOO method improves the model convergence. A lower contrastive loss is achieved using dynamic weighting which translates to an average improvement of $~0.7$ points over the equal weighting setting in zero-shot transfer performance (see Table \ref{tab:moo}).  

\begin{table*}[t!]
\centering
\footnotesize
\scalebox{0.86}{
\begin{tabular}{lccccccccccccc}
\hline
\ \textbf{Target $\rightarrow$} &  \multicolumn{3}{c}{\textbf{WikiQA}} & \multicolumn{3}{c}{\textbf{WikiPassageQA}} & \multicolumn{3}{c}{\textbf{InsuranceQA}} & \multicolumn{3}{c}{\textbf{YahooQA}} & \\
\textbf{Weighting Method} & MAP & MRR & P@1 & MAP & MRR & P@1  & MAP & MRR & P@1  & MAP & MRR & P@1 & \textbf{Avg MAP} \\
\hline
\multicolumn{14}{c}{\textbf{BERT}} \\
\hline
Equal Weighting   
& \underline{84.6} & \underline{86.1} & \underline{75.7} 
& 58.6 & 67.1 & 51.9
& 14.0 & 17.7 & 9.7
& 52.4 & 52.3 & 30.9 & 52.4 \\
Dynamic Weighting  
& \underline{84.6} & \underline{86.0} & \underline{75.8} 
& 58.0 & 66.5 & 50.7 
& 13.6 & 17.2 & 8.7 
& 56.2 & 56.2 & 35.4 & \textbf{53.1} \\
\hline
\end{tabular}
}
\caption{Dynamic weighting vs equal weighting.} 
\label{tab:moo}
\end{table*}

%% file: acl2021.bbl
\begin{thebibliography}{53}
\expandafter\ifx\csname natexlab\endcsname\relax\def\natexlab#1{#1}\fi

\bibitem[{Bhatia et~al.(2020)Bhatia, Liu, Arumae, Pourdamghani, Deshpande,
  Snively, Mona, Wise, Price, Ramaswamy, Ma, Nallapati, Huang, Xiang, and
  Kass-Hout}]{Bhatia2020}
Parminder Bhatia, Lan Liu, Kristjan Arumae, Nima Pourdamghani, Suyog Deshpande,
  Ben Snively, Mona Mona, Colby Wise, George Price, Shyam Ramaswamy, Xiaofei
  Ma, Ramesh Nallapati, Zhiheng Huang, Bing Xiang, and Taha Kass-Hout. 2020.
\newblock \href {http://arxiv.org/abs/2007.09186} {{AWS CORD-19 Search: A
  Neural Search Engine for COVID-19 Literature}}.

\bibitem[{Chechik et~al.(2010)Chechik, Sharma, Shalit, and
  Bengio}]{Chechik2010}
Gal Chechik, Varun Sharma, Uri Shalit, and Samy Bengio. 2010.
\newblock {Large scale online learning of image similarity through ranking}.
\newblock \emph{Journal of Machine Learning Research}, 11:1109--1135.

\bibitem[{Cheng et~al.(2020)Cheng, Jiang, and Macherey}]{Cheng2019}
Yong Cheng, Lu~Jiang, and Wolfgang Macherey. 2020.
\newblock \href {https://doi.org/10.18653/v1/p19-1425} {{Robust neural machine
  translation with doubly adversarial inputs}}.
\newblock In \emph{ACL 2019 - 57th Annual Meeting of the Association for
  Computational Linguistics, Proceedings of the Conference}, 2018, pages
  4324--4333.

\bibitem[{Cohen et~al.(2018)Cohen, Yang, and Croft}]{Cohen2018}
Daniel Cohen, Liu Yang, and W~Bruce Croft. 2018.
\newblock \href {http://arxiv.org/abs/arXiv:1805.03797v1} {{WikiPassageQA: A
  Benchmark Collection for Research on Non-factoid Answer Passage Retrieval}}.
\newblock In \emph{SIGIR}, pages 1--4.

\bibitem[{Devlin et~al.(2018)Devlin, Chang, Lee, and Toutanova}]{Devlin2018}
Jacob Devlin, Ming-Wei Chang, Kenton Lee, and Kristina Toutanova. 2018.
\newblock \href {http://arxiv.org/abs/1810.04805} {{BERT: Pre-training of Deep
  Bidirectional Transformers for Language Understanding}}.

\bibitem[{{dos Santos} et~al.(2020){dos Santos}, Ma, Nallapati, Huang, and
  Xiang}]{NogueiradosSantos2020}
Cicero~Nogueira {dos Santos}, Xiaofei Ma, Ramesh Nallapati, Zhiheng Huang, and
  Bing Xiang. 2020.
\newblock \href {https://doi.org/10.18653/v1/2020.emnlp-main.134} {{Beyond
  [CLS] through Ranking by Generation}}.
\newblock In \emph{Proceedings of the 2020 Conference on Empirical Methods in
  Natural Language Processing (EMNLP)}, pages 1722--1727, Stroudsburg, PA, USA.
  Association for Computational Linguistics.

\bibitem[{{Emma Zhang} et~al.(2019){Emma Zhang}, Sheng, Alhazmi, and
  Li}]{EmmaZhang2019}
Wei {Emma Zhang}, Quan~Z Sheng, Ahoud Alhazmi, and Chenliang Li. 2019.
\newblock \href {https://doi.org/10.1145/nnnnnnn.nnnnnnn} {{Adversarial Attacks
  on Deep Learning Models in Natural Language Processing: A Survey}}.
\newblock 1(1):40.

\bibitem[{Feng et~al.(2016)Feng, Xiang, Glass, Wang, and Zhou}]{Feng2016}
Minwei Feng, Bing Xiang, Michael~R. Glass, Lidan Wang, and Bowen Zhou. 2016.
\newblock \href {https://doi.org/10.1109/ASRU.2015.7404872} {{Applying deep
  learning to answer selection: A study and an open task}}.
\newblock In \emph{2015 IEEE Workshop on Automatic Speech Recognition and
  Understanding, ASRU 2015 - Proceedings}, pages 813--820. Institute of
  Electrical and Electronics Engineers Inc.

\bibitem[{Gan and Ng(2020)}]{Gan2020}
Wee~Chung Gan and Hwee~Tou Ng. 2020.
\newblock \href {https://doi.org/10.18653/v1/p19-1610} {{Improving the
  robustness of question answering systems to question paraphrasing}}.
\newblock In \emph{ACL 2019 - 57th Annual Meeting of the Association for
  Computational Linguistics, Proceedings of the Conference}, pages 6065--6075.

\bibitem[{Gillick et~al.(2019)Gillick, Kulkarni, Lansing, Presta, Baldridge,
  Ie, and Garcia-Olano}]{gillick-etal-2019-learning}
Daniel Gillick, Sayali Kulkarni, Larry Lansing, Alessandro Presta, Jason
  Baldridge, Eugene Ie, and Diego Garcia-Olano. 2019.
\newblock \href {https://www.aclweb.org/anthology/K19-1049} {Learning dense
  representations for entity retrieval}.
\newblock In \emph{Proceedings of the 23rd Conference on Computational Natural
  Language Learning (CoNLL)}, pages 528--537.

\bibitem[{Gunel et~al.(2020)Gunel, Du, Conneau, and Stoyanov}]{Gunel2020}
Beliz Gunel, Jingfei Du, Alexis Conneau, and Ves Stoyanov. 2020.
\newblock \href {http://arxiv.org/abs/2011.01403} {{Supervised Contrastive
  Learning for Pre-trained Language Model Fine-tuning}}.

\bibitem[{Hendrycks et~al.(2019)Hendrycks, Mazeika, Kadavath, and
  Song}]{Hendrycks2019}
Dan Hendrycks, Mantas Mazeika, Saurav Kadavath, and Dawn Song. 2019.
\newblock \href {http://arxiv.org/abs/1906.12340v2} {{Using self-supervised
  learning can improve model robustness and uncertainty}}.
\newblock In \emph{Advances in Neural Information Processing Systems},
  volume~32.

\bibitem[{Hermans et~al.(2017)Hermans, Beyer, and Leibe}]{Hermans2017}
Alexander Hermans, Lucas Beyer, and Bastian Leibe. 2017.
\newblock \href {http://arxiv.org/abs/1703.07737} {{In defense of the triplet
  loss for person re-identification}}.

\bibitem[{Hjelm et~al.(2018)Hjelm, Fedorov, Lavoie-Marchildon, Grewal, Bachman,
  Trischler, and Bengio}]{Hjelm2018}
R~Devon Hjelm, Alex Fedorov, Samuel Lavoie-Marchildon, Karan Grewal, Phil
  Bachman, Adam Trischler, and Yoshua Bengio. 2018.
\newblock \href {http://arxiv.org/abs/1808.06670} {{Learning deep
  representations by mutual information estimation and maximization}}.
\newblock pages 1--24.

\bibitem[{Jin et~al.(2020)Jin, Jin, Zhou, and Szolovits}]{Jin2020}
Di~Jin, Zhijing Jin, Joey~Tianyi Zhou, and Peter Szolovits. 2020.
\newblock \href {https://doi.org/10.1609/aaai.v34i05.6311} {{Is BERT Really
  Robust? A Strong Baseline for Natural Language Attack on Text Classification
  and Entailment}}.
\newblock \emph{Proceedings of the AAAI Conference on Artificial Intelligence},
  34(05):8018--8025.

\bibitem[{Jin et~al.(2004)Jin, Okabe, and Sendhoff}]{Jin2004}
Yaochu Jin, Tatsuya Okabe, and Bernhard Sendhoff. 2004.
\newblock \href {https://doi.org/10.1109/cec.2004.1330830} {{Neural network
  regularization and ensembling using multi-objective evolutionary
  algorithms}}.
\newblock In \emph{Proceedings of the 2004 Congress on Evolutionary
  Computation, CEC2004}, volume~1, pages 1--8.

\bibitem[{Jin et~al.(2001)Jin, Olhofer, and Sendhoff}]{Jin2001}
Yaochu Jin, Markus Olhofer, and Bernhard Sendhoff. 2001.
\newblock \href
  {http://citeseerx.ist.psu.edu/viewdoc/summary?doi=10.1.1.27.9439} {{Dynamic
  Weighted Aggregation for Evolutionary Multi-Objective Optimization:
  {\{}W{\}}hy Does It Work and How?}}
\newblock In \emph{Proceedings of the Genetic and Evolutionary Computation
  Conference GECCO}, pages 1042--1049.

\bibitem[{Karpukhin et~al.(2020)Karpukhin, Oguz, Min, Lewis, Wu, Edunov, Chen,
  and Yih}]{karpukhin-etal-2020-dense}
Vladimir Karpukhin, Barlas Oguz, Sewon Min, Patrick Lewis, Ledell Wu, Sergey
  Edunov, Danqi Chen, and Wen-tau Yih. 2020.
\newblock \href {https://www.aclweb.org/anthology/2020.emnlp-main.550} {Dense
  passage retrieval for open-domain question answering}.
\newblock In \emph{Proceedings of the 2020 Conference on Empirical Methods in
  Natural Language Processing (EMNLP)}, pages 6769--6781.

\bibitem[{Kaushik et~al.(2019)Kaushik, Hovy, and Lipton}]{Kaushik2019}
Divyansh Kaushik, Eduard Hovy, and Zachary~C Lipton. 2019.
\newblock \href {http://arxiv.org/abs/1909.12434} {{Learning the Difference
  that Makes a Difference with Counterfactually-Augmented Data}}.

\bibitem[{Khashabi et~al.(2020)Khashabi, Khot, and Sabharwal}]{Khashabi2020}
Daniel Khashabi, Tushar Khot, and Ashish Sabharwal. 2020.
\newblock \href {http://arxiv.org/abs/2004.04849v1} {{Natural Perturbation for
  Robust Question Answering}}.

\bibitem[{Khosla et~al.(2020)Khosla, Teterwak, Wang, Sarna, Tian, Isola,
  Maschinot, Krishnan, and Liu}]{Khosla2020}
Prannay Khosla, Piotr Teterwak, Chen Wang, Aaron Sarna, Yonglong Tian, Phillip
  Isola, Aaron Maschinot, Dilip Krishnan, and Ce~Liu. 2020.
\newblock \href {http://arxiv.org/abs/2004.11362} {{Supervised contrastive
  learning}}.

\bibitem[{Le-Khac et~al.(2020)Le-Khac, Healy, and Smeaton}]{Le-Khac2020}
Phuc~H Le-Khac, Graham Healy, and Alan~F Smeaton. 2020.
\newblock \href {https://doi.org/10.1109/ACCESS.2020.3031549} {{Contrastive
  Representation Learning: A Framework and Review}}.
\newblock \emph{IEEE Access}, 8:193907--193934.

\bibitem[{Lewis et~al.()Lewis, Liu, Goyal, Ghazvininejad, Mohamed, Levy,
  Stoyanov, and Zettlemoyer}]{Lewis}
Mike Lewis, Yinhan Liu, Naman Goyal, Marjan Ghazvininejad, Abdelrahman Mohamed,
  Omer Levy, Ves Stoyanov, and Luke Zettlemoyer.
\newblock \href {http://arxiv.org/abs/1910.13461v1} {{BART: Denoising
  Sequence-to-Sequence Pre-training for Natural Language Generation,
  Translation, and Comprehension}}.
\newblock Technical report.

\bibitem[{Li et~al.(2019)Li, Yu, Chen, and Li}]{Li2019}
Dongfang Li, Yifei Yu, Qingcai Chen, and Xinyu Li. 2019.
\newblock \href {http://arxiv.org/abs/1905.07588} {{BERTSel: Answer selection
  with pre-trained models}}.

\bibitem[{Lu et~al.(2020)Lu, Abrego, Ma, Ni, and Yang}]{lu2020neural}
Jing Lu, Gustavo~Hernandez Abrego, Ji~Ma, Jianmo Ni, and Yinfei Yang. 2020.
\newblock \href {http://arxiv.org/abs/2010.12523} {Neural passage retrieval
  with improved negative contrast}.

\bibitem[{Ma et~al.(2019)Ma, Xu, Wang, Nallapati, and Xiang}]{Ma2019}
Xiaofei Ma, Peng Xu, Zhiguo Wang, Ramesh Nallapati, and Bing Xiang. 2019.
\newblock \href {https://doi.org/10.18653/v1/d19-6109} {{Domain Adaptation with
  BERT-based Domain Classification and Data Selection}}.
\newblock pages 76--83. Association for Computational Linguistics (ACL).

\bibitem[{Min et~al.(2020)Min, McCoy, Das, Pitler, and Linzen}]{Min2020}
Junghyun Min, R.~Thomas McCoy, Dipanjan Das, Emily Pitler, and Tal Linzen.
  2020.
\newblock \href {https://doi.org/10.18653/v1/2020.acl-main.212} {{Syntactic
  Data Augmentation Increases Robustness to Inference Heuristics}}.
\newblock pages 2339--2352.

\bibitem[{Musgrave et~al.(2020)Musgrave, Belongie, and
  Lim}]{musgrave2020pytorch}
Kevin Musgrave, Serge Belongie, and Ser-Nam Lim. 2020.
\newblock \href {http://arxiv.org/abs/2008.09164} {Pytorch metric learning}.

\bibitem[{Pan and Yang(2010)}]{QiangYang2010}
Sinno~Jialin Pan and Qiang Yang. 2010.
\newblock \href {https://doi.org/10.1109/TKDE.2009.191} {{A survey on transfer
  learning}}.
\newblock \emph{IEEE Transactions on Knowledge and Data Engineering},
  22(10):1345--1359.

\bibitem[{Radford et~al.(2021)Radford, Kim, Hallacy, Ramesh, Goh, Agarwal,
  Sastry, Askell, Mishkin, Clark, Kr{\"u}ger, and Sutskever}]{radford2021}
A.~Radford, J.~W. Kim, Chris Hallacy, Aditya Ramesh, G.~Goh, Sandhini Agarwal,
  Girish Sastry, Amanda Askell, Pamela Mishkin, J.~Clark, G.~Kr{\"u}ger, and
  Ilya Sutskever. 2021.
\newblock Learning transferable visual models from natural language
  supervision.

\bibitem[{Raffel et~al.(2019)Raffel, Shazeer, Roberts, Lee, Narang, Matena,
  Zhou, Li, and Liu}]{Raffel2019}
Colin Raffel, Noam Shazeer, Adam Roberts, Katherine Lee, Sharan Narang, Michael
  Matena, Yanqi Zhou, Wei Li, and Peter~J. Liu. 2019.
\newblock \href {http://arxiv.org/abs/1910.10683} {{Exploring the Limits of
  Transfer Learning with a Unified Text-to-Text Transformer}}.

\bibitem[{Ribeiro et~al.(2018)Ribeiro, Singh, and Guestrin}]{Ribeiro2018}
Marco~Tulio Ribeiro, Sameer Singh, and Carlos Guestrin. 2018.
\newblock \href {https://doi.org/10.18653/v1/p18-1079} {{Semantically
  equivalent adversarial rules for debugging NLP models}}.
\newblock In \emph{ACL 2018 - 56th Annual Meeting of the Association for
  Computational Linguistics, Proceedings of the Conference (Long Papers)},
  volume~1, pages 856--865.

\bibitem[{Ribeiro et~al.(2020)Ribeiro, Wu, Guestrin, and Singh}]{Ribeiro2020}
Marco~Tulio Ribeiro, Tongshuang Wu, Carlos Guestrin, and Sameer Singh. 2020.
\newblock \href {http://arxiv.org/abs/2005.04118} {{Beyond Accuracy: Behavioral
  Testing of NLP models with CheckList}}.
\newblock pages 4902--4912.

\bibitem[{dos Santos et~al.(2016)dos Santos, Tan, Xiang, and
  Zhou}]{santosTXZ16}
C{\'{\i}}cero~Nogueira dos Santos, Ming Tan, Bing Xiang, and Bowen Zhou. 2016.
\newblock \href {http://arxiv.org/abs/1602.03609} {Attentive pooling networks}.
\newblock \emph{CoRR}, abs/1602.03609.

\bibitem[{Shah et~al.(2019)Shah, Chen, Rohrbach, and Parikh}]{Shah2019}
Meet Shah, Xinlei Chen, Marcus Rohrbach, and Devi Parikh. 2019.
\newblock \href {https://doi.org/10.1109/CVPR.2019.00681} {{Cycle-consistency
  for robust visual question answering}}.
\newblock In \emph{Proceedings of the IEEE Computer Society Conference on
  Computer Vision and Pattern Recognition}, volume 2019-June, pages 6642--6651.

\bibitem[{Si et~al.(2020)Si, Yang, Cui, Ma, Liu, and Wang}]{Si2020}
Chenglei Si, Ziqing Yang, Yiming Cui, Wentao Ma, Ting Liu, and Shijin Wang.
  2020.
\newblock \href {http://arxiv.org/abs/2004.14004} {{Benchmarking robustness of
  machine reading comprehension models}}.

\bibitem[{Sun et~al.(2020)Sun, Hashimoto, Yin, Asai, Li, Yu, and
  Xiong}]{Sun2020}
Lichao Sun, Kazuma Hashimoto, Wenpeng Yin, Akari Asai, Jia Li, Philip Yu, and
  Caiming Xiong. 2020.
\newblock \href {http://arxiv.org/abs/2003.04985} {{Adv-BERT: BERT is not
  robust on misspellings! Generating nature adversarial samples on BERT}}.

\bibitem[{Tay et~al.(2017)Tay, Phan, Tuan, and Hui}]{Tay2017}
Yi~Tay, Minh~C. Phan, Luu~Anh Tuan, and Siu~Cheung Hui. 2017.
\newblock \href {https://doi.org/10.1145/3077136.3080790} {{Learning to rank
  question answer pairs with holographic dual LSTM architecture}}.
\newblock In \emph{SIGIR 2017 - Proceedings of the 40th International ACM SIGIR
  Conference on Research and Development in Information Retrieval}, pages
  695--704. Association for Computing Machinery, Inc.

\bibitem[{{Van Den Oord} et~al.(2018){Van Den Oord}, Li, and
  Vinyals}]{VanDenOord2018}
Aaron {Van Den Oord}, Yazhe Li, and Oriol Vinyals. 2018.
\newblock \href {http://arxiv.org/abs/1807.03748} {{Representation learning
  with contrastive predictive coding}}.

\bibitem[{Wang et~al.(2017)Wang, Zhou, Wen, Liu, and Lin}]{Wang2017}
Jian Wang, Feng Zhou, Shilei Wen, Xiao Liu, and Yuanqing Lin. 2017.
\newblock \href {https://doi.org/10.1109/ICCV.2017.283} {{Deep Metric Learning
  with Angular Loss}}.
\newblock In \emph{Proceedings of the IEEE International Conference on Computer
  Vision}, volume 2017-Octob, pages 2612--2620. Institute of Electrical and
  Electronics Engineers Inc.

\bibitem[{Wang and Bansal(2018)}]{Wang2018}
Yicheng Wang and Mohit Bansal. 2018.
\newblock \href {https://doi.org/10.18653/v1/n18-2091} {{Robust Machine
  Comprehension Models via Adversarial Training}}.
\newblock pages 575--581.

\bibitem[{Wang et~al.(2020)Wang, Ng, Ma, Nallapati, and Xiang}]{Wang2020}
Zhiguo Wang, Patrick Ng, Xiaofei Ma, Ramesh Nallapati, and Bing Xiang. 2020.
\newblock \href {https://doi.org/10.18653/v1/d19-1599} {{Multi-passage BERT: A
  globally normalized BERT model for open-domain question answering}}.
\newblock In \emph{EMNLP-IJCNLP 2019 - 2019 Conference on Empirical Methods in
  Natural Language Processing and 9th International Joint Conference on Natural
  Language Processing, Proceedings of the Conference}, pages 5878--5882.
  Association for Computational Linguistics.

\bibitem[{Wei and Zou(2019)}]{Wei2019}
Jason~W. Wei and Kai Zou. 2019.
\newblock \href {http://arxiv.org/abs/1901.11196} {{EDA: Easy Data Augmentation
  Techniques for Boosting Performance on Text Classification Tasks}}.

\bibitem[{Weinberger et~al.()Weinberger, Blitzer, and Saul}]{Weinberger}
Kilian~Q Weinberger, John Blitzer, and Lawrence~K Saul.
\newblock {Distance Metric Learning for Large Margin Nearest Neighbor
  Classification}.
\newblock Technical report.

\bibitem[{Wolf et~al.(2019)Wolf, Debut, Sanh, Chaumond, Delangue, Moi, Cistac,
  Rault, Louf, Funtowicz, and Brew}]{wolf2019HuggingFacesTS}
Thomas Wolf, Lysandre Debut, Victor Sanh, Julien Chaumond, Clement Delangue,
  Anthony Moi, Pierric Cistac, Tim Rault, R'emi Louf, Morgan Funtowicz, and
  Jamie Brew. 2019.
\newblock Huggingface's transformers: State-of-the-art natural language
  processing.
\newblock \emph{ArXiv}, abs/1910.03771.

\bibitem[{Wu et~al.(2020)Wu, Wang, Gu, Khabsa, Sun, and Ma}]{Wu2020}
Zhuofeng Wu, Sinong Wang, Jiatao Gu, Madian Khabsa, Fei Sun, and Hao Ma. 2020.
\newblock \href {http://arxiv.org/abs/2012.15466} {{CLEAR: Contrastive Learning
  for Sentence Representation}}.

\bibitem[{Xiong et~al.(2020)Xiong, Xiong, Li, Tang, Liu, Bennett, Ahmed, and
  Overwijk}]{xiong2020approximate}
Lee Xiong, Chenyan Xiong, Ye~Li, Kwok-Fung Tang, Jialin Liu, Paul Bennett,
  Junaid Ahmed, and Arnold Overwijk. 2020.
\newblock \href {http://arxiv.org/abs/2007.00808} {Approximate nearest neighbor
  negative contrastive learning for dense text retrieval}.

\bibitem[{Xu et~al.(2019)Xu, Ma, Nallapati, and Xiang}]{Xu2019}
Peng Xu, Xiaofei Ma, Ramesh Nallapati, and Bing Xiang. 2019.
\newblock \href {http://arxiv.org/abs/1905.05910} {{Passage ranking with weak
  supervision}}.

\bibitem[{Yang et~al.(2015)Yang, Yih, and Meek}]{Yang2015}
Yi~Yang, Wen-Tau Yih, and Christopher Meek. 2015.
\newblock {WIKIQA: A Challenge Dataset for Open-Domain Question Answering}.
\newblock \emph{Proceedings of EMNLP 2015}, (September 2015):2013--2018.

\bibitem[{Yin et~al.(2020)Yin, Long, Meng, and Chang}]{Yin2020}
Fan Yin, Quanyu Long, Tao Meng, and Kai-Wei Chang. 2020.
\newblock \href {https://doi.org/10.18653/v1/2020.acl-main.310} {{On the
  Robustness of Language Encoders against Grammatical Errors}}.
\newblock pages 3386--3403.

\bibitem[{Zhang et~al.(2019)Zhang, Liu, Long, and Jordan}]{Zhang2019}
Yuchen Zhang, Tianle Liu, Mingsheng Long, and Michael~I Jordan. 2019.
\newblock \href {http://arxiv.org/abs/1904.05801} {{Bridging theory and
  algorithm for domain adaptation}}.
\newblock In \emph{36th International Conference on Machine Learning, ICML
  2019}, volume 2019-June, pages 12805--12823.

\bibitem[{Zhou et~al.(2020)Zhou, Huang, and Zhu}]{Zhou2020}
Mantong Zhou, Minlie Huang, and Xiaoyan Zhu. 2020.
\newblock \href {https://doi.org/10.1109/TASLP.2020.3016132} {{Robust Reading
  Comprehension with Linguistic Constraints via Posterior Regularization}}.
\newblock \emph{IEEE/ACM Transactions on Audio Speech and Language Processing},
  28:2500--2510.

\bibitem[{Zhou et~al.()Zhou, Zeng, Zhou, Anastasopoulos, and Neubig}]{Zhou}
Shuyan Zhou, Xiangkai Zeng, Yingqi Zhou, Antonios Anastasopoulos, and Graham
  Neubig.
\newblock \href {http://www.statmt.org/wmt15/} {{Improving Robustness of Neural
  Machine Translation with Multi-task Learning}}.
\newblock Technical Report~1.

\end{thebibliography}
